# Microwave Sensing of Andreev Bound States in a Gate-Defined Superconducting Quantum Point Contact

Vivek Chidambaram[1,2], Anders Kringhøj[2], Lucas Casparis[2,3], Ferdinand Kuemmeth[2], Tiantian Wang[4], Candice Thomas[4], Sergei Gronin[5,4], Geoffrey C. Gardner[5,4], Zhengyi Cui[6], Chenlu Liu[6], Kristof Moors[7], Michael J. Manfra[4,5], Karl D. Petersson[2,3], and Malcolm R. Connolly[2,6]

[1]*Semiconductor Physics Group, Cavendish Laboratory, University of Cambridge, JJ Thomson Avenue, Cambridge CB3 0HE, United Kingdom*

[2]*Center for Quantum Devices, Niels Bohr Institute, University of Copenhagen, Universitetsparken 5, 2100 Copenhagen, Denmark*

[3] *Microsoft Quantum Lab-Copenhagen, Niels Bohr Institute, University of Copenhagen, 2100 Copenhagen, Denmark*

[4] *Department of Physics and Astronomy, Purdue University, West Lafayette, IN, USA*

[5]*Microsoft Quantum Purdue, West Lafayette, IN, USA*

[6]*Blackett Laboratory, Imperial College London, South Kensington Campus, London SW7 2AZ, United Kingdom*

[7]*Institute for Semiconductor Nanoelectronics, Peter Grünberg Institute 9, Forschungszentrum Jülich, Germany*

**Keywords**

Superconductors, Semiconductors, Andreev Bound States, Microwave Spectroscopy, Circuit QED

**Abstract**

We use a superconducting microresonator as a cavity to sense absorption of microwaves by a superconducting quantum point contact defined by surface gates over a proximitized two-dimensional electron gas. Renormalization of the cavity frequency with phase difference across the point contact is consistent with coupling to Andreev bound states. Near π phase difference, we observe random fluctuations in absorption with gate voltage, related to quantum interference-induced modulations in the electron transmission. Close to pinch-off, we identify features consistent with the presence of a single Andreev bound state and describe the Andreev-cavity interaction using a dispersive Jaynes-Cummings model. By fitting the weak Andreev-cavity coupling, we extract ~GHz decoherence consistent with charge noise and the transmission dispersion associated with a localized state.

The flow of supercurrent across a Josephson junction (JJ) is described by Andreev bound states (ABSs), coherent superpositions of electrons and holes that transport Cooper pairs by Andreev reflection at the junction interfaces.[1] By setting the Josephson inductance $L_J$, ABSs play a central role in superconducting transmon qubits[2] and also form a two-level basis for spin-based quasiparticle qubits.[3] In the short-junction limit, ABSs are spatially confined to the JJ with energy $E_A = \pm\Delta\sqrt{1 - \tau_n \sin^2(\delta/2)}$, where Δ is the superconducting gap, δ is the phase across the JJ, and $\tau_n$ the transmission probability of the *n*-th electron channel in the JJ material.[4] Conventional qubits based on aluminium either host a fixed high number of low-τ ABSs in oxide JJs, yielding the familiar Josephson energy $E_J = E_0 \cos(\delta)$, or a few mechanically tuneable high-τ ABSs used for quasiparticle qubits in atomic point contacts.[5] Rapid improvement in the growth of hybrid superconductor-semiconductor interfaces over the last decade has unlocked semiconductor JJs that can span these operating regimes, using local capacitively-coupled gate electrodes to tune the number[6,7] and $\tau_n$ distribution[8] of ABSs.

Gatemons, the voltage-controlled variant of a transmon,[9] have so far been realised with semiconductor nanowires,[10,11] carbon,[12,13,14,15] and topological insulator[16] JJs. Single channels in InAs nanowire gatemons were recently associated with suppressed relaxation[17] and charge dispersion.[18,19] Splitting of microwave transitions in quasiparticle qubits[20] due to strong spin-orbit interaction[21] has enabled spin-to-supercurrent conversion[22] and paves the way towards time-domain analysis of Majorana bound states.[23] Qubit states can be probed by coupling the dipole transition to a transmission line via a microwave cavity resonator. Capacitive coupling is used for transmons and inductive coupling for quasiparticle qubits, generating a state-dependent shift of the cavity frequency described by circuit quantum electrodynamics (cQED).[24] Proximitized two-dimensional electron gases (2DEGs) are an attractive scalable and versatile platform for realising voltage-controlled JJs, with ABS confinement potentials tailored by the outline of lithographically-patterned surface gates. Gatemons with 2DEG JJs have microsecond lifetimes, limited by microwave losses in the InP substrate.[25] Here we establish the 2DEG variant for quasiparticle qubits by studying ABSs in a gate-defined superconducting quantum point contact (SQPC), shown schematically in Fig. 1(a). Our 2DEG comprises a trilayer InGaAs/InAs/InGaAs stack grown on a buffered InP substrate. A 50 nm-thick epitaxial Al capping layer proximitizes the 2DEG via Andreev reflection while preserving high ($\sim 1$ m$^2$V$^{-1}$s$^{-1}$) charge carrier mobilities.[26] ABSs are lithographically confined by laterally etching a mesa with width $W_1 \sim 4$ μm down to the InP buffer layer and removing a $L \sim 100$ nm strip of Al. To form a SQPC the ABSs are further electrostatically confined to a narrow constriction, by depleting carriers from regions under the split-gate electrodes, which are spaced by width $W_2 \sim 100$ nm.

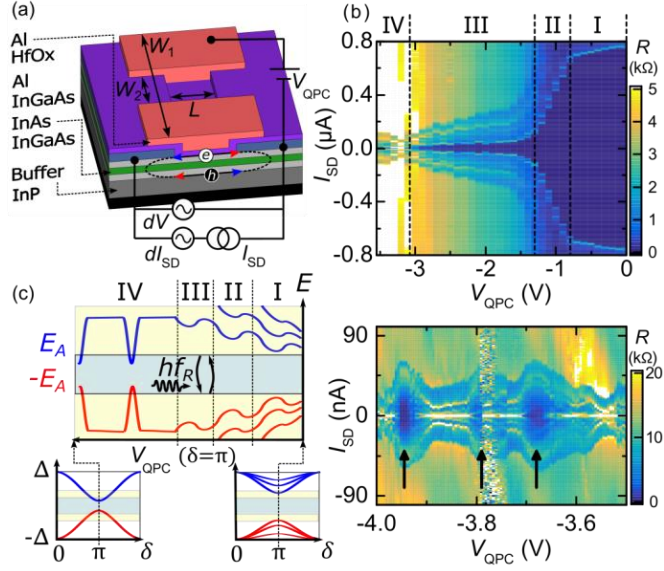

**Figure 1.** (a) Stack structure of an InAs 2DEG with split-gates (red) used to define a SQPC. ABSs comprise right/left- (red) and left/right- (blue) propagating electron/hole pairs Andreev reflected at the interface. (b) Differential resistance $R \sim dV/dI_{SD}$ of a SQPC as a function of $V_{QPC}$ and $I_{SD}$. Regions: open (I), definition (II), depletion (III), and tunnelling (IV). Lower panel shows $R$ over a narrower range of $I_{SD}$ in region IV, with $I_c$ resonances approaching ~45 nA indicated by arrows. (c) Schematic ABS energy spectrum of the junction reconstructed as a function of phase difference $\delta = \theta_2 - \theta_1$ (lower panels) and $V_{QPC}$ at $\delta = \pi$ (upper). In each region, the energy detuning between bare cavity photons ($f_R$) and ABS transitions ($f_A = 2E_A/h$) determines the effective shift in the observed cavity frequency.

Fig. 1(b) shows a plot of the differential resistance $R \sim dV/dI_{SD}$ as a function of gate voltage $V_{QPC}$ and bias current $I_{SD}$ of a typical SQPC. $R$ was determined by conventional AC lock-in measurements in a four-terminal configuration, using an external 1 MΩ resistor to implement current-biased measurements. A dissipationless state ($R = 0$) is maintained up to a maximum critical current $I_c$. In principle $I_c = \sum_N I_c^{(n)}$, where $I_c^{(n)} \sim e\Delta/\hbar(1 - \sqrt{1-\tau_n})$ is carried by each of the $N$ ABS channels, each with transmission probability $\tau_n$. In a ballistic SQPC with transparent interfaces ($\tau_n \sim 1$), steps in $I_c(V_{QPC})$ with size $I_0 \sim e\Delta/\hbar \sim 50$ nA are the anticipated counterpart of conductance quantization $G_0 \sim 2e^2/h$ in a normal QPC.[4] Our devices typically do not exhibit steps, but have four regions with distinctive behaviour, delineated by the dashed lines in Fig. 1 (b), and referred to as open (region I), definition (region II), depletion (region III), and tunnelling (region IV). In region I, $I_c$ drops only slightly from ∼0.8 μA to ∼0.7 μA with increasing negative $V_{QPC}$ as the number of channels reduces beneath the split gates. Using a carrier density $n_{2D} \sim 3 \times 10^{16}$ m$^{-2}$ and mobility $\mu \sim 0.5$ m$^2$V$^{-1}$s$^{-1}$, determined separately from Hall measurements on the same wafer, yields a Fermi energy $E_F = n_{2D}\pi\hbar^2/m^* \sim 0.3$ eV, assuming an effective mass of $m^* = 0.023 m_e$. The resulting Fermi wavelength $\lambda_F \sim 14$ nm and channel number $N = 2W_1/\pi\sqrt{2m^*E_F}/\hbar \sim 1000$. This is reasonably consistent with the measured above gap (normal state) resistance $R_N \sim 100$ Ω expected for a disordered JJ with $N \sim \Delta/\delta E \sim R_K/R_N \sim 250$ channels with high average $\tau_n$, where $R_K$ is the resistance quantum and $\delta E$ is the energy-level spacing. The observed $I_c \sim 1$ μA, however, corresponds to an order of magnitude lower number of channels ($N \sim 20$), a feature also reported in an earlier study on wide JJs.[27] We speculate that channels with higher momentum parallel to the interface have path lengths $\Lambda > L$ that exceed the superconducting coherence length $\xi$, and thus do not contribute to the transfer of Cooper pairs. Using the conductivity $\sigma = \nu e^2 D$ and $\nu = m^*/\pi\hbar^2$ for the constant 2D density of states, we obtain $D \sim 0.07\ m^2 s^{-1}$ for the diffusion coefficient, $\tau_{sc} = m^*\mu/e \sim 2.5$ ps for the scattering time, and mean free path $l_{mfp} = \sqrt{D\tau_{sc}} \sim 70$ nm, placing our JJs between the diffusive and ballistic regimes. $\xi^*$ is between the clean ($\sim \hbar v_F/\pi\Delta^* \sim 1.4$ μm) and dirty ($\sim \sqrt{\xi_0 l_{mfp}} \sim 300$ nm) limits, where $v_F \sim 1.2 \times 10^6$ is the Fermi velocity, and $\Delta^* \sim 180$ μeV is the induced gap measured by bias spectroscopy. In either case our devices satisfy $L < \xi^*$ and a critical current contribution of $I_0$ per channel is not expected in region I since $W_2 \sim W_1$.[28] During definition (region II), $I_c$ drops over a $\Delta V_{QPC} \sim 0.5$ V interval as channels contributing to $I_c$ either side of the SQPC are closed. Once the SQPC is defined, $I_c \sim I_0$ and reduces gradually over a $\Delta V_{QPC} \sim 1.5$ V as $W_1$ decreases (region III). The reduced $dI_c/dV_{QPC}$ in this region results from the reduced capacitive coupling from the split gates to the constriction.

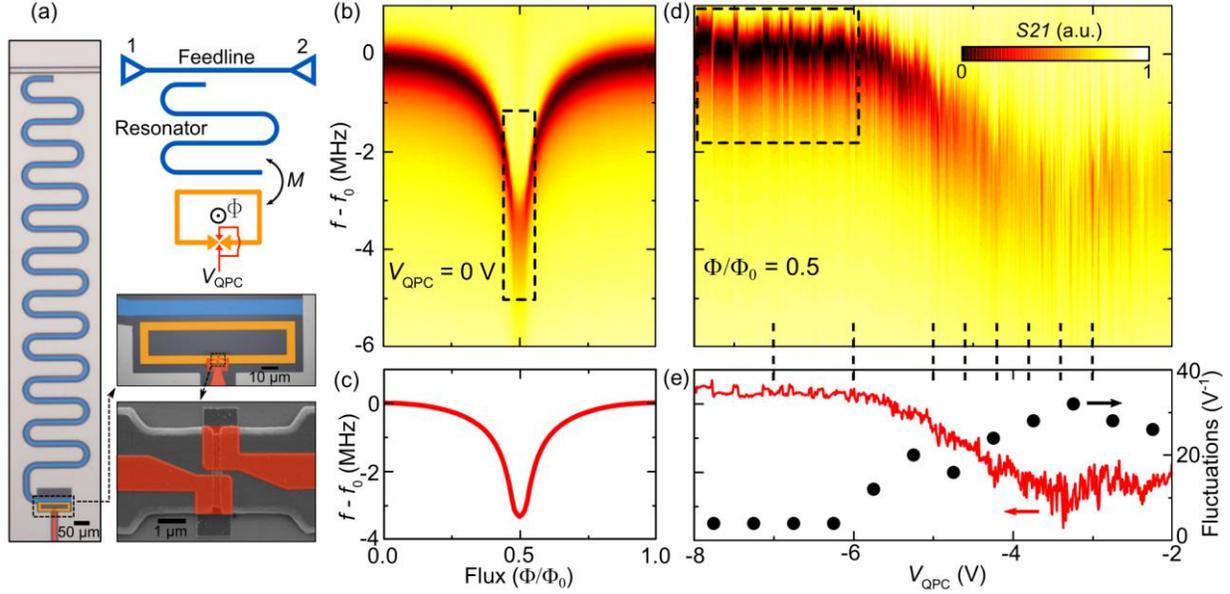

**Figure 2.** (a) Optical and scanning electron micrographs of the device. A serpentine microwave cavity (blue) is inductively coupled to a loop of Al/InAs (orange) interrupted by a SQPC (red). (b) Feedline transmission coefficient S21 as a function of feedline frequency $f$ and flux $\Phi$. $f_0 \sim 6.163$ GHz is the observed resonance frequency at zero flux at zero gate voltage. (c) Resonant frequency shift extracted from (b). (d) S21 as a function of frequency and $V_{QPC}$ at $\Phi/\Phi_0 = 0.5$. (e) Number of cavity fluctuations per volt (black points) and resonant frequency (red line) extracted from (d).

In region IV [Fig. 1(b)], $R_N$ fluctuates around $h/2e^2$ and $I_c$ displays $\Delta V_{QPC} \sim 50$ mV-wide resonances. We attribute this to the opening and closing of a single highly-transmitting channel, highlighted by arrows in the lower panel of Fig. 1(b). Similar behaviour observed in nanowire gatemons was correlated with suppressed charge dispersion[18,19] due to resonant tunnelling (RT). To link this behaviour to the microwave properties, Fig. 1(c) shows a sketch of the microwave energy gap $2E_A(V_{QPC})$ for these four regions, reconstructed by assuming $\tau$ is tuned by $V_{QPC}$ and using $I_c \sim dE_A/d\delta$. The degeneracy between counterpropagating electron-hole pairs at $\delta = \pi$ is lifted due to backscattering, opening a gap $2\Delta(\sqrt{1-\tau_n})$. Microwaves with energy $hf_A = 2E_A$, typically $\sim 5-10$ GHz for high-$\tau_n$ channels,[5] should be absorbed by the JJ, effectively exciting a quasiparticle and reversing the direction of supercurrent. In this picture $I_c$ resonances in region IV correspond to peaks (dips) in transmission (Andreev transition frequency), suitable for gate-defined quasiparticle qubits.

To probe this microwave spectrum experimentally, we report data from a different device comprising a SQPC in a superconducting loop inductively coupled to a NbTiN microwave cavity [Fig. 2(a)]. As described previous studies,[29,30] the cavity frequency depends on the current-phase relation (CPR) and the occupancy of the ABS ground ($-E_A$) and excited ($+E_A$) states. Below we focus on two features of the observed cavity behaviour, namely the overall modulation $\Delta f_R$ of the resonance frequency with the applied magnetic flux through the loop, and finer shifts with $V_{QPC}$. Figure 2(b) shows the dependence of the amplitude of the microwave transmission (S21) on the reduced flux $\Phi' = \Phi/\Phi_0$ (applied via a single coil mounted to the sample enclosure) and readout difference frequency $f - f_0$, where $\Phi_0$ is the flux quantum and $f_0 = 6.163$ GHz is the resonant frequency of the bare cavity identified as the sharp reduction in S21 at $\Phi' = 0$. In the open regime, $f_R$ decreases rapidly and loses visibility towards $\Phi' = 0.5$. In Fig. 2(d) we fix the flux at $\Phi' = 0.5$ and plot S21 as a function of $V_{QPC}$. In the range $-4 < V_{QPC} < 0$ we observe rapid $\sim 1$ MHz shifts in cavity frequency, fluctuating $\sim 30$ V$^{-1}$ around the original $f_R(0.5)$ [see Fig. 2(e)]. At intermediate voltages ($-6 < V_{QPC} < -4$) the fluctuations persist with similar density while the average cavity frequency shifts upwards closer to $f_R(0)$ and the quality factor increases from ~2000 to ~7000. For $V_{QPC} < -6$ V, $f_R$ is roughly constant and the number of fluctuations drops to ~5 V$^{-1}$. The overall trend in $f_R(V_{QPC})$ qualitatively follows the same evolution shown in regions I – III in Fig. 1(b), implying the cavity push relates to the total $I_c$. This is consistent with recent work[31] demonstrating that in the adiabatic regime ($f_R \ll f_A$) the cavity frequency is renormalised by the Josephson inductance $L_J = (\Phi_0/2\pi)^2(\partial^2 E_A^{(n)}/\partial \delta^2)^{-1}$ of each channel shown schematically in Fig. 1(c), where $\delta = 2\pi\Phi'$. Over the full frequency range, the total cavity shift $\Delta f_R = \sum_N \Delta f_R^{(n)}$, where $\Delta f_R^{(n)}$ has both first- (adiabatic) and second- (dispersive) order contributions. In Fig. 2(c) we plot the flux modulation of the cavity resonance frequency $\Delta f_R = f_R(\Phi') - f_0$ and find good agreement with $N\sim6$ channels (see Appendix for details of the model and an explicit comparison with the experimental data). This is lower than the $N\sim20$ from the transport measurements shown in Fig. 1 but consistent with the as-fabricated $I_c$ seen in similar devices.

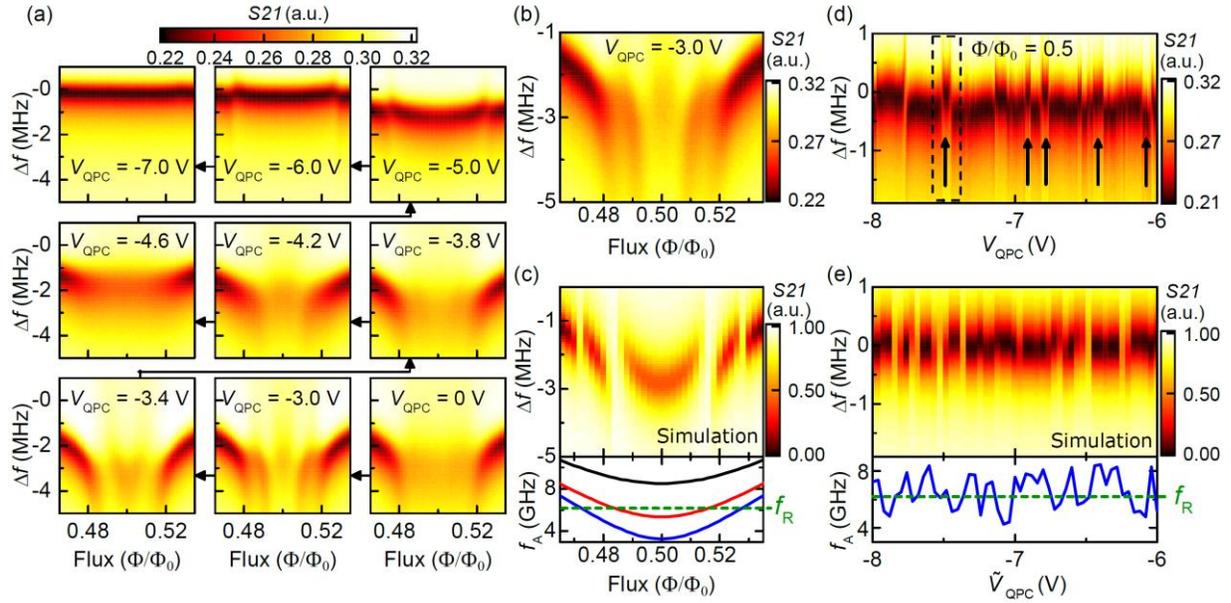

**Figure 3.** (a) S21 as a function of $f$ and $\Phi'$ for $V_{QPC}$ values indicated by the dashed lines in Fig. 2(d). Measured and simulated S21 as a function of (b,c) $\Phi'$ and (d,e) $V_{QPC}$ and readout difference frequency. Lower panels show the transition frequency for 3 Andreev channels used in the simulations. $\tilde{V}_{QPC}$ is the simulation sweep parameter scaled to the same range as $V_{QPC}$.

Having observed the correlation between $\Delta f_R$ and $I_c$, it is perhaps suprising that the fluctuations observed in $\Delta f_R(V_{QPC}, 0.5)$ are typically absent from $I_c(V_{QPC}, 0)$ in region I. To elucidate this we measure $S21(\Phi', f)$ for $V_{QPC}$ values shown by dashed lines in Fig. 2(d). The series of plots in Fig. 3(a) show the decrease in overall push $\Delta f_R(V_{QPC}, 0.5)$ with increasing negative $V_{QPC}$, and also reveals a variety of avoided crossings symmetric about $\Phi' = 0.5$. Avoided crossings are a sign of virtual cavity-ABS photon exchange, inducing a push on the cavity frequency described by a Jaynes-Cummings (JC) interaction. In the JC picture, avoided crossings result from a divergence in the cavity push $\chi_{ij}^{(JC)} = g_{ij}^2/2\pi(f_R - f_{ij})$, where $g_{ij}$ is the coupling rate, when the detuning from the $i \to j$ ABS transition approaches $f_{ij} \sim f_R$.[28] Generally $f_{ij}$ comprises spin-orbit and subband kinetic energy contributions from different parity manifolds.[21] For transitions within the same manifold, $f_{ij} = f_A^{(n)}(V_{QPC}, \pi \pm \Delta\delta_n)$, channels with sufficient $\tau_n$ meet this condition at different phase offset $\Delta\delta_n$, giving rise to the different patterns of crossings in Fig. 3(a).

In order to capture these features quantitatively we simulate the coupled ABS-cavity system using a master equation for the JC model (see Appendix).[32] Note the dispersive JC shift $\sum \chi_{ij}^{(JC)}$ is valid in the range of flux shown in Fig. 3(a). Figure 3(b) shows the raw $S21(\Phi', f)$ data at $V_{QPC} = -3$ V and corresponding simulation in Fig. 3(c). The discrete crossings and overall cavity shift $\sim -2$ MHz are well reproduced assuming $N = 3$ spin-degenerate channels, consistent with modelling of Fig. 2(b). We also confirmed that the model reproduces the dependence on $V_{QPC}$ at $\Phi' = 0.5$ [Figs. 3(d) and (e)] using the same model parameters. Note that the $\tilde{V}_{QPC}$ of the simulation is a randomly generated value of $\tau_n$, related to the underlying statistics of universal conductance fluctuations (UCFs).[30] Using $\Delta E_F = \alpha_1 \Delta V_{QPC} \pi \hbar^2 / m^* e^2$, where $\alpha_1 = \varepsilon_{HfO}/d$ is the capacitance to the 2DEG, $d = 20$ nm, and $\varepsilon_{HfO} = 11$, and equating the number of avoided crossings $1/\Delta V_{QPC} \sim 30$ $V^{-1}$, where $\Delta V_{QPC}$ is the correlation voltage, yields $\Delta E_F \sim 10$ meV. This is in agreement with the intermediate limit ($L > l_{mfp}$),[33] where fluctuations of $\sim I_0$ are expected when $\Delta E_F \sim \Delta(\sqrt{\xi^* l_{mfp}}/L) \sim 15$ meV (for $\xi^* \sim 1$ μm). In this picture the factor of $\sim 3$ increase in the correlation voltage [Fig. 2(e)] in region IV is due to the weaker capacitive coupling between the split gate and the constriction. UCFs therefore appear in the cavity push as it only couples to the lowest modes, while in transport they are averaged out from region I in $I_c$ [Fig. 1(b)].

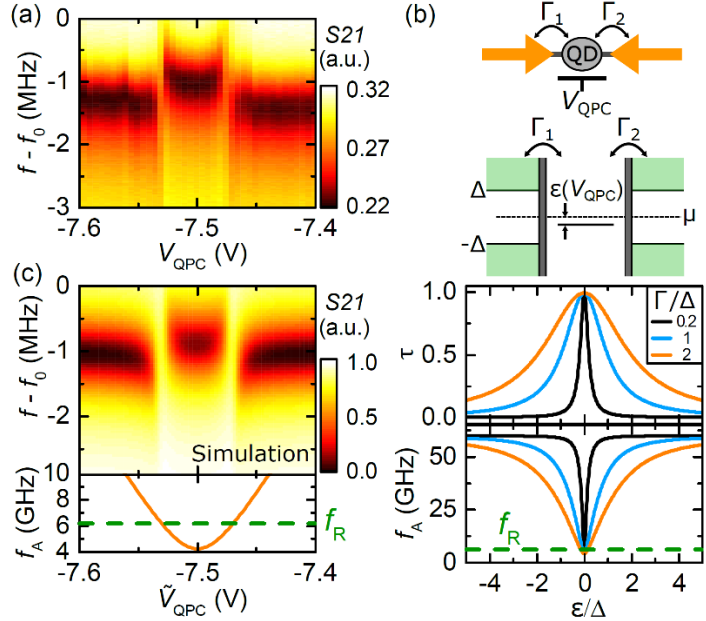

**Figure 4.** (a) S21 measured as a function of frequency and $V_{QPC}$ at the paired crossings highlighted by the dashed box in Fig. 3(d). (b) Schematic of a quantum dot formed in the junction, coupled to the leads with tunneling rates $\Gamma_1$ and $\Gamma_2$. The energy level sketch (middle panel) shows a quantum dot state with energy detuning $\varepsilon$ from the chemical potential μ modulated by $V_{QPC}$. (Lower panel) Plot of Breit-Wigner transmission ($\Gamma_1 = \Gamma_2 = \Gamma$) and ABS frequency. (c) Simulated S21 as a function of $\tilde{V}_{QPC}$ using the $f_A(\varepsilon)$ shown in lower inset.

A notable feature below $V_{QPC} \sim -6$ V is the presence of paired crossings [see arrows in Fig. 3(d)]. We focus on the pair highlighted by the dashed box in Fig. 3(d) and shown in detail in Fig. 4(a). Paired crossings are naturally explained by peaks in $\tau(V_{QPC})$, as illustrated in region IV of Fig. 1(c). Resonant passage at the the conduction band edge is possible along periodically spaced impurity configurations,[34] or a quantum dot[35] formed by a potential valley exceeding the lowest subband spacing, $E_2 - E_1 \sim 5$ meV, where $E_n \sim n^2 \pi^2 \hbar^2 / 2m^* W_1$. Following recent work,[18,19] we implement the RT model shown in Fig. 4(b), where a discrete level in a quantum dot is tunnel coupled to left and right leads with rate $\Gamma_1$ and $\Gamma_2$, respectively. The resulting transmission as a function of detuning $\tau(\varepsilon)$ is shown in Fig. 4(b) for different $\Gamma/\Delta$ together with the dip in $f_A$. For symmetric barriers, RT enables perfect transmission regardless of the details of the backscattering potential. The simulation shows good agreement with the experimental data [Fig. 4(c)], with $\Gamma/\Delta \sim 1$, consistent with results from simulations of nanowire devices ($\Gamma = 60 - 72$ GHz),[18] and maximum $\tau \sim 0.995$. Note the fact that transmission is not unity is due to some residual asymmetry in tunnel rates $\Gamma_1$ and $\Gamma_2$. The $I_c$ resonances in region IV of Fig. 1(b) also have a $\Delta V_{QPC} \sim 50$ mV width, and are thus naturally accommodated within the RT picture.

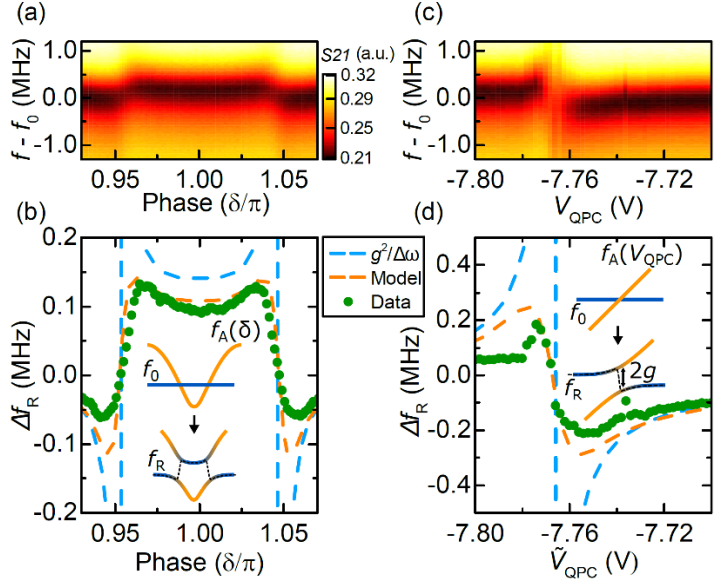

**Figure 5**. (a) S21 measured as a function of $\Delta f$ and $\delta$ in the single-channel regime ($V_{QPC} \sim -6$ V). (b) Shift in cavity frequency extracted from the S21 data in (a) and plotted as a function of $\delta$ (green points). Fit to the experimental data using the J-C model (orange dashed) and dispersive shift (blue dashed). (c) S21 measured as a function of frequency and $V_{QPC}$ at $\delta = \pi$. (d) Shift in cavity frequency shift with $V_{QPC}$ (green points), J-C model (orange dashed), and dispersive shift (blue dashed). Insets in (b) and (d) illustrate the ABS-cavity hybridisation.

Finally, since the ABS-cavity interaction is sensitive to quasiparticle dynamics, we can estimate the lifetime $T_\varphi$ via the decoherence rate $\gamma = 1/T_\varphi$ in the JC model. We fix $V_{QPC}$ and measure $S21(\Phi', f)$ where a single ABS transition hybridises with the cavity [Fig. 5(a)] and extract the cavity frequency to compare with the JC result and the simple dispersive shift $g^2/2\pi\Delta f$ [Fig. 5(b)]. Figure 5(c) and (d) also shows the equivalent plots for the avoided crossing when the ABS transition is tuned with $V_{QPC}$. The JC model, unlike the dispersive shift, shows good agreement with both data sets using $\kappa \approx 1$ MHz (estimates from a 6.163 GHz resonator with internal $Q \sim 7000$) and $g/2\pi \sim 16$ MHz, corresponding to a $M \sim 10$ pH, yielding $\gamma/2\pi \sim 1$ GHz. The fact $g \ll \gamma$ accounts for the weak coupling and absence of clear Rabi splitting, with similar behaviour observed in multiple devices. Note that $M$ is $\sim 3$ times lower than extracted from the experimental data in Fig. 2 taken in the many-channel regime. This could be due to an underestimate of the channel number in the many-channel regime, which would imply lower $M$, or to using the JC approximation in the single-mode regime. The $T_\varphi \sim 1$ ns is shorter by several orders of magnitude than atomic point contacts[5] and recent work with InAs nanowire JJs.[19] We speculate that the main mechanism for inelastic quasiparticle relaxation are emission or absorption of phonons[36] and electromagnetic coupling to the environment.[37] Charge noise in 2DEGs is expected to generate root-mean-squared (RMS) fluctuations of $\langle V_{QPC} \rangle_{RMS} \sim 0.5$ mV.[25] The slope $df_A/dV_{QPC} \sim 120$ GHz/V [Fig. 4(c)] yields $T_2^* \sim 4$ ns,[38] comparable to our measured $T_\varphi$. Nanowire gatemons with RT showed off-resonant $df_Q/dV_G \sim 500$ GHz/V,[19] but the lower noise $\langle V_G \rangle_{RMS} \sim 10$ μV[35] and suppressed charge dispersion improves $T_2^* \sim 50$ ns.[19] Note that odd-parity states in which a single quasiparticle occupies the ABS would limit coherence on the ~μs timescale and thus strongly limit coherent manipulation. We anticipate further improvements by using thicker buffer layers to increase the carrier mobility, decreasing the gate capacitance, and reducing the maximum frequency in longer channels.

In summary we have used a superconducting microresonator to probe the microwave response of a 2DEG SQPC as a function of phase difference and carrier density. A monotonic shift in cavity frequency as a function of phase is consistent with coupling to ABSs channels. In the dispersive regime close to $\pi$ phase across the JJ, we observed random shifts in the cavity frequency as a function of gate voltage related to quantum interference-induced fluctuations in transmission. We reproduced the phase and gate-voltage dependence using a JC model. In the single-channel regime we observe paired crossings as a function of gate voltage, suggesting the presence of resonant tunnelling in a quantum dot. The absence of Rabi splitting is due to the $\sim\mathrm{GHz}$ decoherence induced by gate noise and level dispersion. With optimisation of materials our study paves the way for quantum control of gate-defined quasiparticle qubits and provides insight into how the gate geometry and material of a JJ relates to qubit performance.

**Acknowledgements**

We acknowledge helpful discussions with Charles Marcus, Thorvald Larsen, Natalie Pearson and Albert Hertel. This work was supported by the European Union's Horizon 2020 research and innovation programme under the Marie Sklodowska-Curie grant agreement No. 750777, the Engineering and Physical Sciences Research Council (EP/L020963/1), and Microsoft Project Q. K. M. acknlowledges the financial support by the Bavarian Ministry of Economic Affairs, Regional Development and Energy within Bavaria's High-Tech Agenda Project "Bausteine für das Quantencomputing auf Basis topologischer Materialien mit experimentellen und theoretischen Ansätzen" (grant allocation no. 07 02/686 58/1/21 1/22 2/23).

**Appendix**

**1. Theoretical model describing the many-channel Andreev-cavity coupling**

We describe the superconducting quantum point contact (SQPC) using the low-energy effective Hamiltonian of the 2DEG:

$$H^{2DEG} = -\frac{\hbar^2}{2m^*}\left(\partial_x^2 + \partial_y^2\right) - i\alpha\left(\partial_x\sigma_y - \partial_y\sigma_x\right) + U(y), \quad (1)$$

where $\alpha$ is the Rashba spin-orbit strength, $\sigma_{x,y}$ are the Pauli spin matrices, and $U(y)$ is the transverse electrostatic potential profile induced by the split gate. We use Kwant[39] to solve (1) on a two-dimensional tight-binding lattice for the electron modes $\psi_e(x, y)$ as a function of the chemical potential $\mu_{QPC}$, which is controlled by the applied split-gate voltage $V_{QPC}$. We consider both parabolic (PB) and hardwall (HW) confining potentials $U_{PB/HW}(y)$ given by[40]:

$$U_{HW}(y) = \begin{cases} \mu_{QPC}, & |y| < \frac{W_y}{2} \\ \infty, & |y| > \frac{W_y}{2} \end{cases}$$

$$U_{PB}(y) = \mu_{QPC} + \Omega_y(\mu_{QPC})y^2$$

In this model, increasing $\mu_{QPC}$ shifts the potential minimum and increases the lateral confinement (given by the width $W_y$ for the HW and the curvature $\Omega_y$ for PB) to capture the effect of increasing $V_{QPC}$ in the device. To determine the energy-phase relationship $E_A(\delta)$ we use the short-junction scattering matrix formalism,[33] $S_A S_N \psi = \psi$, where the mode vector $\psi = (\psi_e, \psi_h)$ describes equally weighted electron and hole states, $S_N$ ($S_A$) is the scattering matrix of normal (Andreev) scattering between electrons $S_e(E, \mathbf{k})$ and holes $S_h(E, \mathbf{k})$. We make use of the standard expression for $S_A$ based on wavefunction matching at the SN interface.[41] Applying $\text{Det}[1 - S_A S_N] = 0$ allows us to solve for $E_A(\delta)/\Delta$ for each channel. Then using $E_A(\delta = \pi)/\Delta = \sqrt{1-\tau}$, we extract the transmission $\tau_i$ of each mode. Based on the maximum junction critical current $\sim 1$ μA we set the chemical potential $\mu_{SC}$ in the superconducting leads such that a maximum of $N = 20$ channels are incident on the SQPC. The calculated transmissions $\{\tau_i\}$ are shown as a function of $\mu_{QPC}$ in Fig. 6(a) and (b) for both HW and PB potentials. Note that the channels are spin-degenerate, so each line represents two channels. As $\mu_{QPC}$ increases, the transmission of each channel drops to near-zero. One or two pairs of channels remain highly transmitting and undergo fluctuations in transmission before finally being backscattered as the QPC is pinched off. To visualise the overall number of highly-transmitting channels the number of channels with $\tau_i > 0.5$ is plotted in 6(c), showing a monotonic decrease with $\mu_{QPC}$ which is qualitatively similar for both HW and PB potentials.

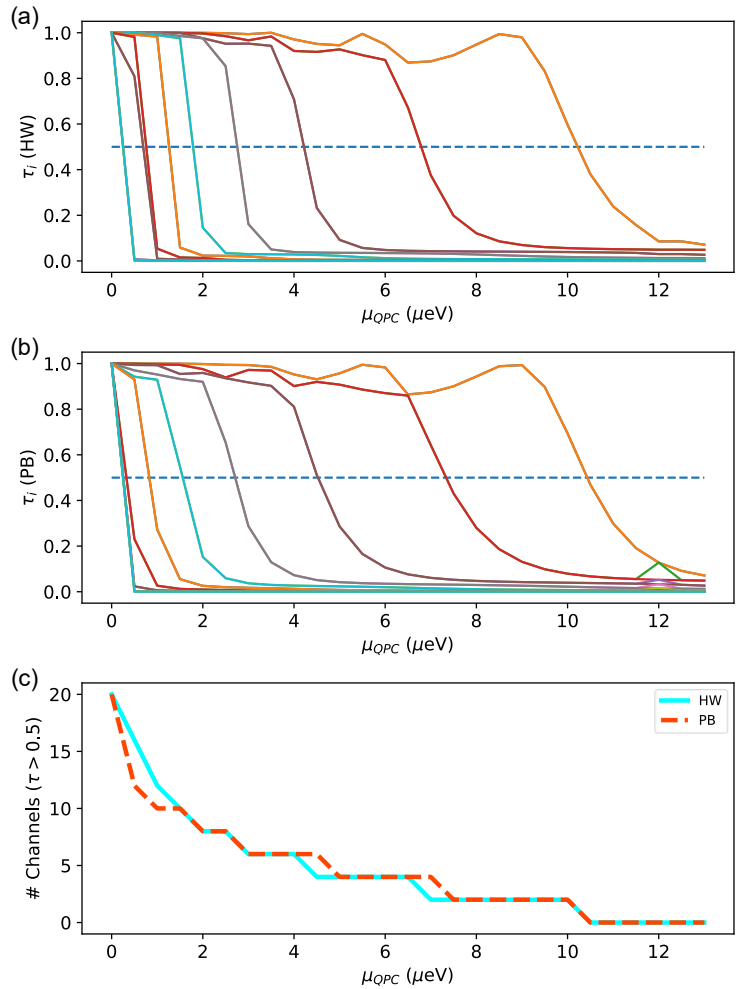

**Fig. 6.** Mode transmissions $\{\tau_i\}$ for the hardwall (HW) potential in (a) and the parabolic (PB) potential in (b) as a function of chemical potential $\mu_{QPC}$. (c) The number of highly-transmitting channels, defined with $\tau_i > 0.5$, showing a progressive decrease with increasing $\mu_{QPC}$ as each channel is backscattered.

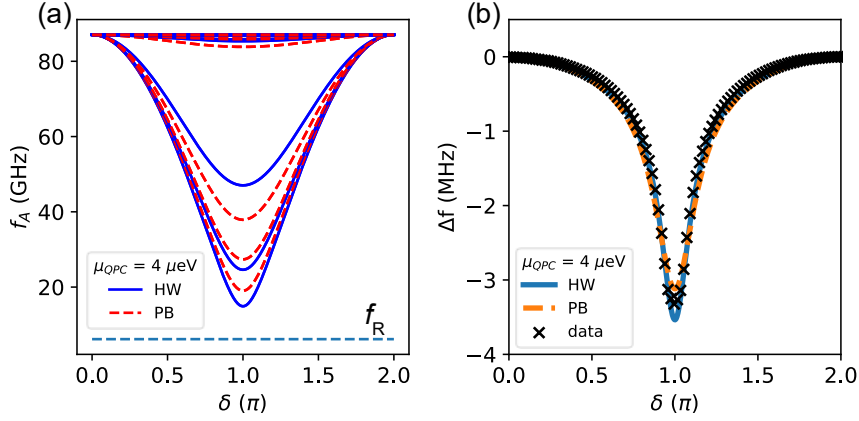

**Fig. 7.** (a) Plot showing the Andreev transition frequency as a function of phase difference for hardwall (HW, blue solid) and parabolic (PB, red dashed) confinement at $\mu_{QPC} = 4\,\mu eV$. (b) Shift in cavity frequency as a function of phase difference for HW (blue solid) and PB (orange dashed), and the experimental data (black crosses).

The Andreev transition frequency $f_A(\delta) = 2E_A(\delta)/h$ for the PB and HW potentials with $\mu_{QPC} = 4\,\mu$eV is shown in Fig. 7(a). We employ the theory for adiabatic and dispersive shifts described in Ref. 31 to calculate the corresponding cavity shift $\Delta f_R(\delta)$ over the full range of large ($f_A \gg f_R$) and small ($f_A \sim f_R$) Andreev-cavity detuning. The mutual inductance $M$ between the cavity and loop determines the coupling strength. The simulated $\Delta f_R(\delta)$ is shown in Fig. 7(b) along with the measured frequency shift data of Fig. 2(c). A good fit is obtained for $\mu_{QPC} = 4\,\mu$eV, with 3 pairs of highly-transmitting channels with $\tau_i^{HW}$ = {0.97, 0.92, 0.71} and $\tau_i^{PB}$ = {0.95, 0.90, 0.81}) and $M = 30$ pH. This is larger than the $M \sim 10$ pH deduced from comparison of the master equation simulation to data in the single-channel regime (Fig. 5). This discrepancy could be due to the precise QPC potential deviating from the PB or HW induced by the split-gate, as this would modify the number of modes and precise $\tau$-distribution in the QPC at a given $V_{QPC}$.

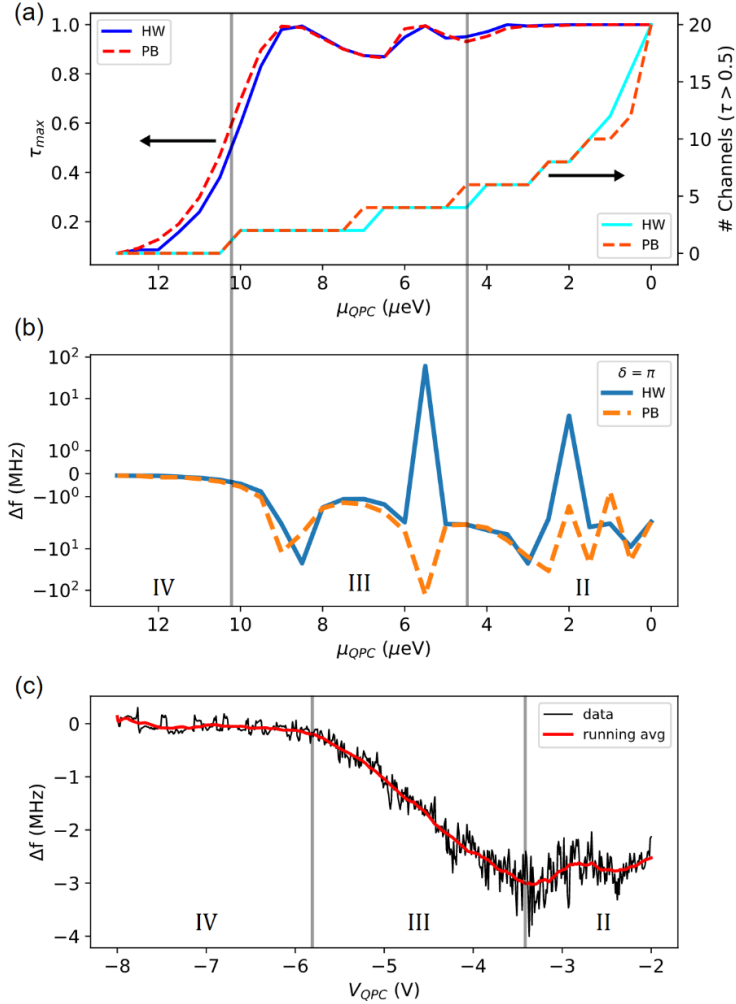

**Fig. 8.** (a) Plot showing the number of transmitting channels as a function of chemical potential for hardwall (HW) and parabolic (PB) confinement. Also shown is the transmission $\tau_{max}$ of the lowest energy mode. (b) Calculated frequency shift of the cavity as a function of chemical potential parameter $\mu_{QPC}$. (c) Experimental frequency shift with gate voltage $V_{QPC}$ along with a running average to show the overall frequency shift excluding fluctuations. Regions II-IV are indicative only and correspond to those indicated in Fig. 1(b).

To understand the measured $\Delta f_R(V_{QPC})$ shown in Fig. 2(e), we plot the evolution of the channel number and the transmission of the lowest energy mode ($\tau_{max}$) as a function of $\mu_{QPC}$ for the PB and HW QPC in Fig. 8(a). As expected, at low $\mu_{QPC}$ the number of channels drops rapidly while the lowest energy mode remains highly transmitting. The corresponding $\Delta f_R(\mu_{QPC}, \pi)$ in Fig. 8(b) shows $\sim -5$ MHz average shift and fluctuates rapidly as each mode experiences a peak in $\tau$ that enables $f_A \approx f_R$. There are a larger number of weaker fluctuations in the experimental data [Fig. 8(c)] due to potential disorder fluctuations and the noise-induced weak coupling analysed in Fig. 5. The overall shift reduces between 8-12 $\mu$eV as the lowest mode is pinched off, consistent with the behaviour for $V_{QPC} < -4$ V in the experimental data.

**2. Description of the few mode Andreev - cavity system**

In Figs. 3-5, numerical simulations are compared to resonator S21 data to understand the dynamics of single- or few-channel ABSs coupled to the resonator. The ABS - cavity Hamiltonian consists of three terms $H = H_R + H_A + H_I$, corresponding to the cavity, Andreev two-level system (TLS), and interaction energies, respectively. The interaction is an inductive coupling term $H_I = M I_R I_A$, where $M$ is the mutual inductance and $I_R$ and $I_A$ are the current operators for the cavity and ABSs respectively. These terms are given by:

$$H_R = \hbar \omega_R \left( a^\dagger a + \frac{1}{2} \right)$$

$$H_A = \frac{\hbar \omega_A}{2} \sigma_z$$

$$H_I = g(a^\dagger + a)\left( \sigma_x + \frac{1}{\sqrt{1-\tau} \tan\left(\frac{\delta}{2}\right)} \sigma_z \right)$$

$$g(\delta, \tau) = \sqrt{z} \frac{E_A(\pi, \tau)}{2} \left( \frac{\Delta}{E_A(\delta, \tau)} - \frac{E_A(\delta, \tau)}{\Delta} \right), \qquad z = \frac{M^2 \omega_R^2}{Z_R R_Q}$$

where $a^+$ and $a$ are creation and annihilation operators acting on the resonator state and $\sigma_i$ are the Pauli operators acting on the even-parity ABS subspace. It can be shown that the interaction term reduces to the standard Jaynes-Cummings (JC) form $H_I \approx g(a^+ \sigma^- + a \sigma^+)$ by going into the interaction picture and making the rotating wave approximation: $|\omega_A - \omega_R| \ll \omega_R$. (Note we do not make this approximation but highlight the connection to the JC Hamiltonian.) We introduce the collapse operators $c_{ops} = \left[\sqrt{\kappa \cdot (1 + n_{th})} \cdot a,\ \sqrt{\kappa n_{th}} \cdot a^\dagger, \sqrt{\gamma} \cdot \sigma^-\right]$ corresponding to cavity relaxation, cavity excitation, and Andreev TLS relaxation where $\kappa$ is the cavity linewidth, $\gamma$ is the Andreev TLS relaxation rate and $n_{th}$ is the average thermal photon number. We implement the above Hamiltonian and collapse operators in QuTiP[42] and use the *qutip.correlation.spectrum* function to calculate the resonator response as a function of phase $\delta$ or transmission $\tau$ and fitted to the experimental frequency response. To perform the 3-channel simulation shown in Fig. 3(c) we extended to a multi-channel version by introducing new Pauli operators for each additional mode and adding Andreev energy terms $H_A$, coupling terms $H_I$ and collapse operators.

[31] Sunghun Park, C. Metzger, L. Tosi, M. F. Goffman, C. Urbina, H. Pothier, and A. Levy Yeyati, *From Adiabatic to Dispersive Readout of Quantum Circuits*, Phys. Rev. Lett. **125**, 077701 (2020).

[32] J.R. Johansson, P.D. Nation, Franco Nori, *QuTiP 2: A Python framework for the dynamics of open quantum systems*, Computer Physics Communications, **184** 4, 1234 (2013).

[33] C. W. J. Beenakker, *Universal Limit of Critical-Current Fluctuations in Mesoscopic Josephson Junctions*, Phys. Rev. Lett. **67** (27), 3836 (1991).

[34] L. G. Aslamazov and M. V. Fistul', *Resonant tunneling in superconductor-semiconductor-superconductor junctions*, Zh. Eksp. Teor. Fiz. **83**, 1170 (1982).

[35] I. A. Devyatov and M. Yu. Kupriyanov, *Resonant Josephson tunneling through S-I-S junctions of arbitrary size*, JETP **85**, 189 (1997).

[36] G. N. Gol'tsman and K. V. Smirnov, *Electron-phonon interaction in a two-dimensional electron gas of semiconductor heterostructures at low temperatures*, JETP **74**, 474 (2001).

[37] D. A. Ivanov & M. V. Feigel'man, *Phonon relaxation of subgap levels in superconducting quantum point contacts*, JETP **68**, 890 (1998).

[38] O. E. Dial, M. D. Shulman, S. P. Harvey, H. Bluhm, V. Umansky, and A. Yacoby, *Charge noise spectroscopy using coherent exchange oscillations in a singlet–triplet qubit*, Phys. Rev. Lett. **110**, 146804 (2013).

[39] C. W. Groth, M. Wimmer, A. R. Akhmerov, X. Waintal, *Kwant: a software package for quantum transport*, New J. Phys. **16**, 063065 (2014).

[40] M. Geier, J. Freudenfeld, J. T. Silva, V. Umansky, D. Reuter, A. D. Wieck, P. W. Brouwer, and S. Ludwig, *Electrostatic potential shape of gate-defined quantum point contacts*, Phys. Rev. B **101,** 165429 (2020).

[41] Doru Sticlet, Bas Nijholt, and Anton Akhmerov, *Robustness of majorana bound states in the short-junction limit*, Phys. Rev. B **95**, 115421 (2017).

[42] J. R. Johansson, P. D. Nation, and F. Nori, *QuTiP 2: A Python Framework for the Dynamics of Open Quantum Systems*, Comput. Phys. Commun. **184**, 1234 (2013).

22